\documentclass[runningheads]{svmult}

\usepackage{makeidx}   
\usepackage{graphicx}  
\usepackage{subeqnar}  
\usepackage{multicol}  
\usepackage{physprbb}  
\makeindex             

\begin{document}

\title*{The RASS-SDSS galaxy cluster survey.}

\toctitle{Focusing of a Parallel Beam to Form a Point
\protect\newline in the Particle Deflection Plane}

\titlerunning{The RASS-SDSS galaxy cluster survey.}

\author{Popesso Paola\inst{1}
\and Bh\"oringer Hans\inst{1}
\and Voges Wolfgang\inst{1}}

\authorrunning{Popesso Paola et al.}

\institute{Max-Planck-Institut fur extraterrestrische Physik, 85748 Garching, Germany}

\maketitle   

\begin{abstract}
Solid observational evidences   indicate  a strong dependence  of  the
galaxy formation and evolution on the environment. In order to study in
particular the  interaction between  the intracluster medium   and the
evolution  of cluster galaxies, we  have  created a large database
 of
clusters of galaxies based on the largest available X-ray and
 optical
surveys: the ROSAT  All Sky Survey (RASS),  and the Sloan  Digital Sky
Survey (SDSS). We  analyzed the correlation  between the total optical
and  the X-ray cluster luminosity.  The resulting correlation of $L_X$
and $L_{op}$  shows a logarithmic slope  of 0.6, a  value close to the
self-similar correlation.  We analysed also the  cluster mass to light
ratio, by finding  a significant  dependence of  $M/L$ on the  cluster
mass with a logarithmic slope ranging from  0.27 in the  i and r bands
to 0.22 in the z band.
\end{abstract}

\section{The RASS-SDSS galaxy cluster catalog.}
For a detailed comparison of the  appearance of cluster of galaxies in
X-ray and in the optical,  we have compiled  a large database of X-ray
and optical properties of  a sample of  clusters based on the  largest
available X-ray  and optical surveys: the ROSAT  All Sky Survey (RASS)
and  the Sloan Digital Sky  Survey  (SDSS).  The X-ray-selected galaxy
clusters of this RASS-SDSS catalog cover  a wide range of masses, from
groups  of
 $10^{12.5}$ $M_{\odot}$ to   massive clusters of $10^{15}$
$M_{\odot}$
 in the redshift range from  0.002 to 0.45.  The RASS-SDSS
sample  comprises  114 clusters.  For  each  system we  have uniformly
determined the X-ray  and optical properties .   For a subsample of 53
clusters we also compiled the temperature  and the iron abundance from
the literature.

\subsection{Correlating X-ray and optical cluster properties}
The cluster   optical luminosity   is  calculated, after   statistical
background subtraction, as the sum of all the  galaxies within the $N$
magnitude bins.   A tight relation  exists between $L_{op}$ and $L_X$.
In the $log(L_X)$-$log(L_{op})$ plane,  the $L_{op}-L_X$ relation is a
line with slope 0.44$\pm$0.03 and with a scatter of 0.3.

Given the $L_X-M$ relation of Reiprich \& Bh\"oringer 2001 and our new
$L_X-L_{op}$  relation  , also  the mass  should increase  faster than
$L_{op}$.    To check this,  we  used the subsample   of clusters with
temperature, and with the $M-T$ relation of Finoguenov  et al. 2001 we
estimated the mass and analysed the $M-L$ and the $M/L-M$ relations (M
and $L_{op}$  are calculated both   within $r_{500}$).  $M/L$  shows a
clear dependence on the  cluster mass with a  slope from 0.27 in the i
and r bands to 0.22 in the z band.

\begin{figure}
 
\parbox{6.3cm}{\resizebox{\hsize}{!}{\includegraphics{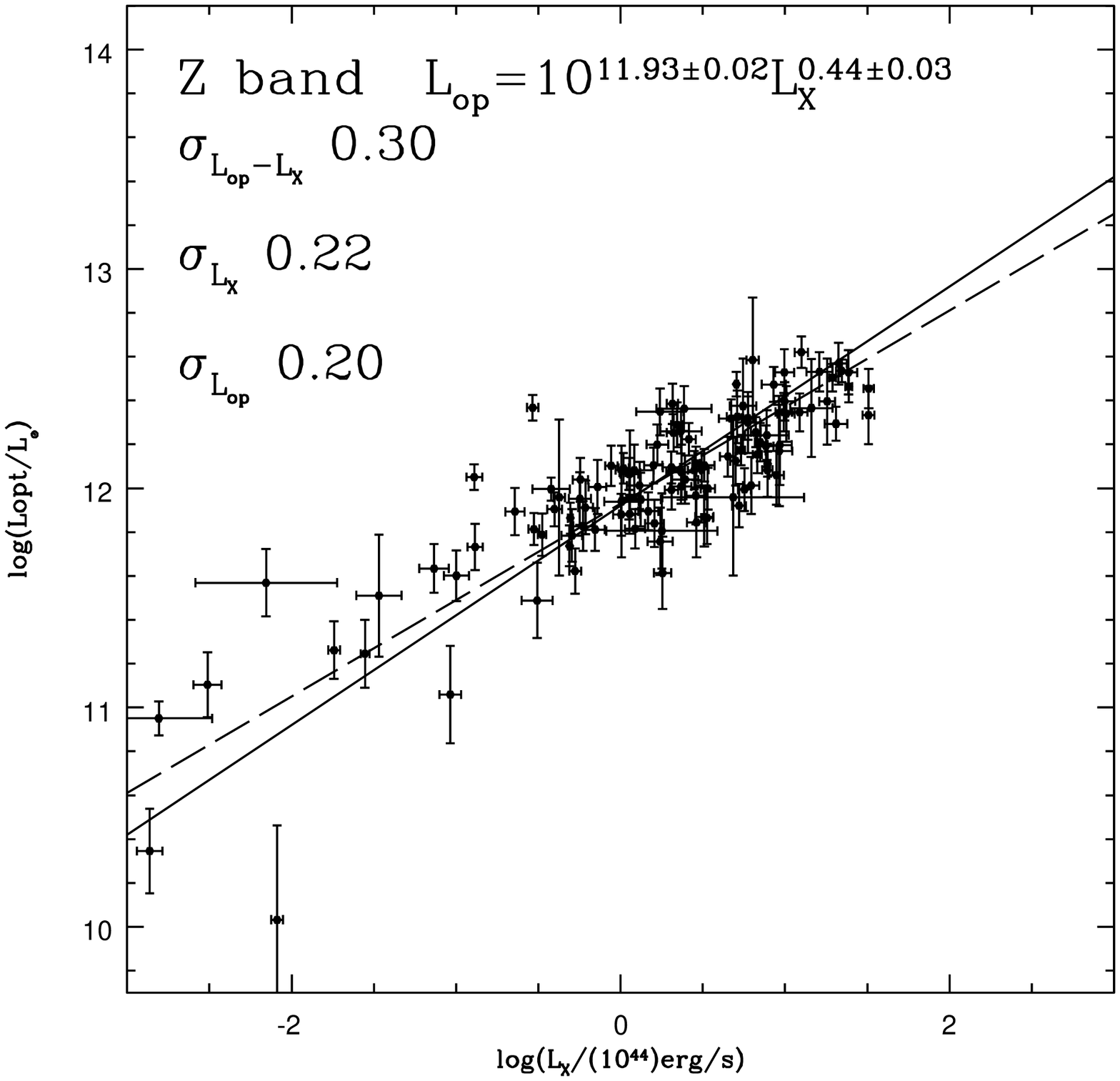}}}
  \hfill
\parbox{6.3cm}{\resizebox{\hsize}{!}{\includegraphics{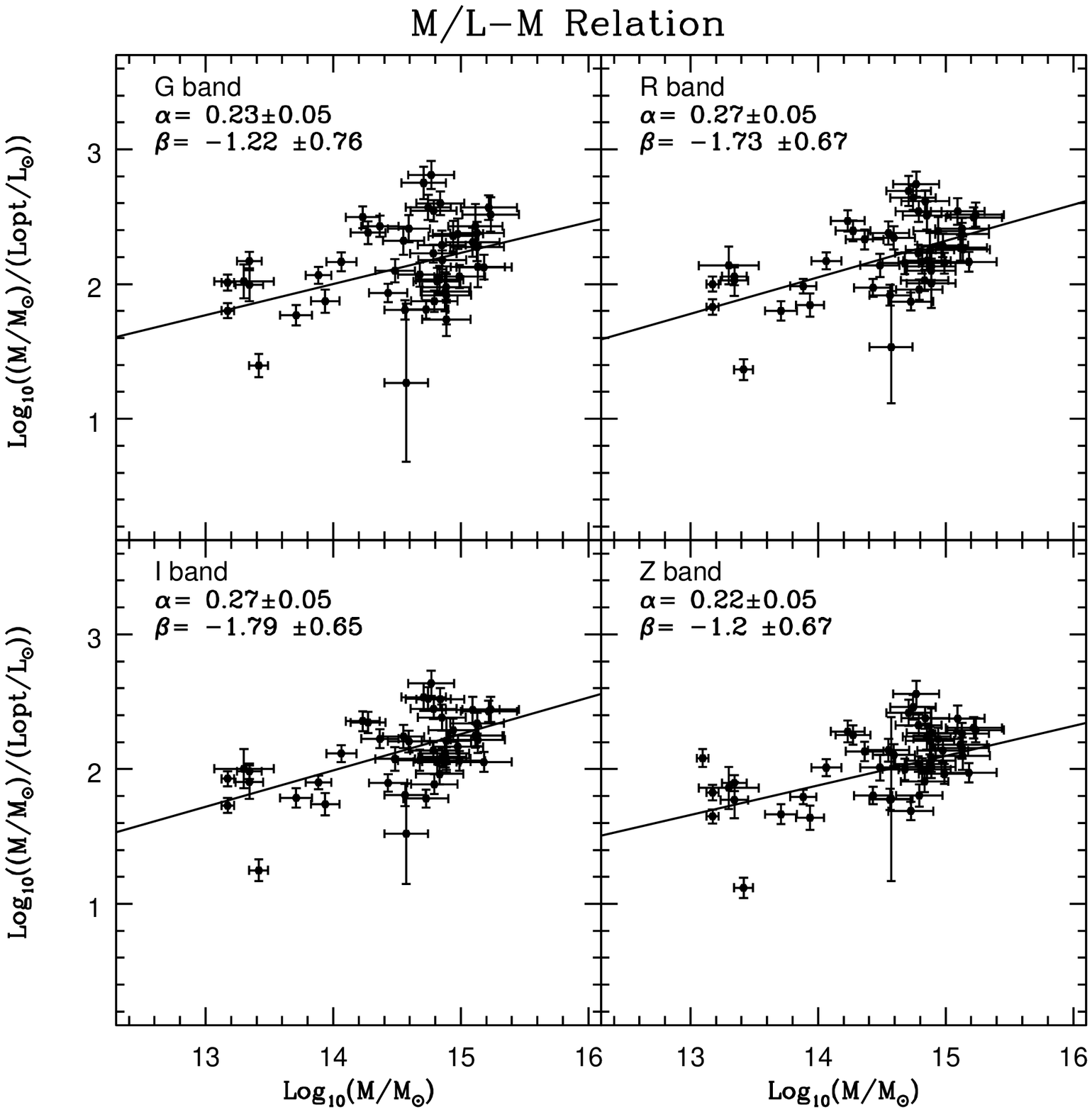}}}
  \\ \\ 
\caption{ A tight relation exists between $L_X$ and  $L_{op}$
(left panel).  The right  panel shows the  cluster $M/L$ as a function
of  the $M_{500}$ mass.  The   mass is estimated from the  temperature
with  the $M_{500}-T_{ICM}$  relation of Finoguenov  et  al. 2001. The
mass increases faster than the  optical luminosity as predicted by our
new  $L_{op}-L_X$ relation.As  a  consequence of the  $M_{500}-L_{op}$
relation,   the $M/L$ ratio is  clearly  an increasing function of the
cluster mass.  This is  in agreement with the  prediction based on our
new $L_X-L_{op}$ relation.}
\label{figure:ecco}
\end{figure}

\vspace{0.5cm}
Funding for the creation and distribution of the SDSS Archive has been
provided  by the   Alfred   P.  Sloan Foundation,    the Participating
Institutions,  the National Aeronautics  and Space Administration, the
National   Science Foundation, the   U.S.   Department of Energy,  the
Japanese Monbukagakusho, and the Max Planck Society. The SDSS Web site
is http://www.sdss.org/.   The  SDSS is managed   by the Astrophysical
Research  Consortium  (ARC) for  the  Participating  Institutions. The
Participating Institutions  are The University  of Chicago,  Fermilab,
the  Institute for Advanced Study, the  Japan Participation Group, The
Johns Hopkins   University,  Los  Alamos   National Laboratory,    the
Max-Planck-Institute for  Astronomy (MPIA), the   Max-Planck-Institute
for Astrophysics (MPA),  New  Mexico State University,  University  of
Pittsburgh, Princeton University, the United States Naval Observatory,
and the University of Washington.

\end{document}